\documentstyle[12pt,epsf]{ioplppt}

\begin{document}

\title{Persistence of Vibrational Modes in a Classical
Two-Dimensional Electron Liquid}

\author{Satoru Muto and Hideo Aoki}

\address{Department of Physics, University of Tokyo, Hongo,
Tokyo 113-0033, Japan}

\begin{abstract}
Vibrational density of states of a classical two-dimensional
electron system obtained with a molecular-dynamics simulation
is shown to have a peak in both solid and liquid phases.
From an exact diagonalisation of the dynamical matrix,
the peak is identified to be vibrational modes having
wavelengths of the order of the electron spacing,
and the result is interpreted as persistent vibrational
modes with short wavelengths in a liquid.
\end{abstract}


\maketitle

It is an interesting question to ask to what extent
vibrational modes remain well-defined in liquid states.
A typical system that is known to have a solid-liquid
transition is the classical two-dimensional electron system
(2DES; electrons interacting with the Coulomb repulsion
in a uniform neutralising positive background), which
has attracted a great deal of experimental and theoretical
interests, especially in the context of melting
in two dimensions (see references cited in \cite{muto-aoki}).
A classical 2DES can be realised on a liquid-helium
surface \cite{grimes-adams}, where electrons obey
classical statistics since the Fermi energy is much
smaller than the thermal energy.
Both experimental and numerical studies, including a recent
one \cite{muto-aoki}, show that the solid-liquid transition
occurs around the plasma parameter
\[
\Gamma \equiv (e^{2}/4\pi\epsilon a)/k_{B}T \simeq 130,
\]
where $a=(\pi n)^{-1/2}$ is the mean electron separation
with $n$ being the density of electrons.

Numerical simulations have been extensively done for
the classical 2DES.
In some of the early studies with the molecular-dynamics (MD)
method, which allows one to calculate dynamical quantities
as well as thermodynamic quantities, an oscillation
is observed in the velocity autocorrelation function
$Z(t) \equiv \left\langle \sum_{i} {\bf v}_{i}(t) \cdot
{\bf v}_{i}(0) \right\rangle / \left\langle \sum_{i}
{\bf v}_{i}^{2}(0) \right\rangle$, whose period is
almost independent of temperature \cite{hansen, kalia}.
That is, the oscillation in $Z(t)$ persists for large $t$
even in the liquid phase.
These authors, however, do not give physical interpretations,
and this has motivated us to look into the problem in more
detail.

We use the canonical MD method developed by
Nos\'{e} \cite{nose} and Hoover \cite{hoover}
for 900 electrons to investigate dynamical properties of
a classical 2DES, while the previous simulations were done
for microcanonical ensembles for smaller systems.
The aspect ratio of the unit cell is taken to be
$L_{y}/L_{x}=2/\sqrt{3}$, which can accommodate a
perfect triangular lattice \cite{bonsall} with $N=4M^{2}$
($M$: an integer) electrons.
We impose periodic boundary conditions and use the Ewald
sum to take care of the long-range nature of the Coulomb
interaction.

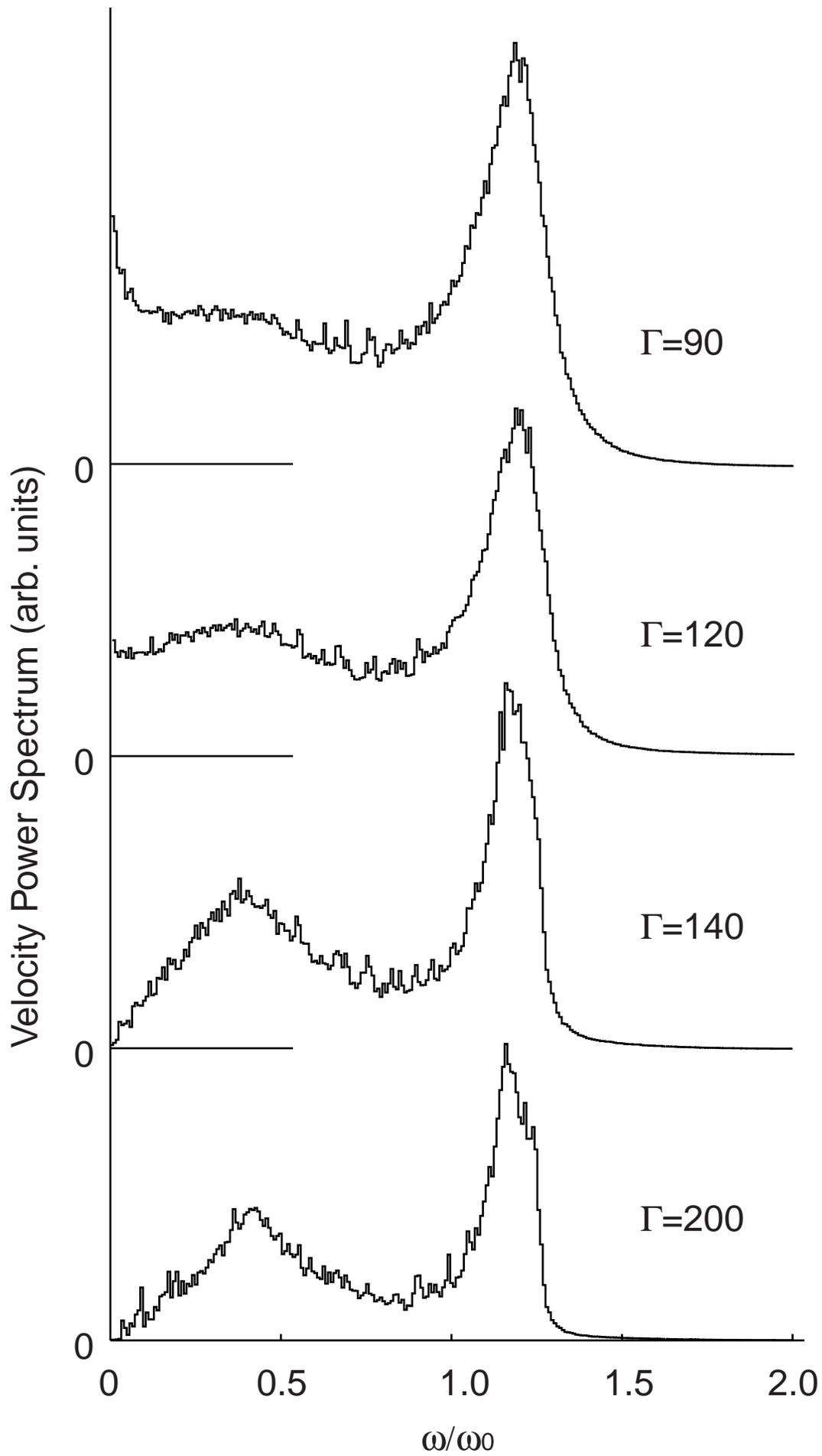
\begin{figure}
\caption{MD result for the velocity power spectrum for
various temperatures ($\propto 1/\Gamma$).
Each spectrum is shown with an offset in the vertical axis.
The result is obtained by heating the system from a solid
to a liquid. The solid-liquid transition occurs at about
$\Gamma = 130$.}
\label{fig:vps}
\end{figure}

First, we present the velocity power spectrum
(vibrational density of states) in figure~\ref{fig:vps},
which corresponds to Fourier transform of the velocity
autocorrelation (Wiener-Khinchin's theorem).
The result shows that the spectrum at zero frequency,
which is proportional to the diffusion constant,
is vanishingly small in the solid phase while finite in the
liquid phase, in agreement with the conventional view that
diffusion distinguishes solids from liquids.
As the temperature increases (i.e., as $\Gamma$ decreases),
the low-frequency components grow in the liquid phase,
indicating larger diffusion at higher temperatures.

Remarkably, we find a peak around $\omega \approx 1.2 \omega_{0}$
that persists when $\Gamma$ is decreased to the liquid regime.
Here the frequency is normalised by $\omega_{0} \equiv
(e^{2}/\sqrt{3}\epsilon m d^{3})^{1/2}$, where
$d=(\sqrt{3}n/2)^{-1/2}$ is the triangular lattice constant.
$\omega_{0}$ is defined in such a way that the longitudinal
(plasma) mode has $\omega (q) = \omega_{0}(qd)^{1/2}$ in
the long wavelength limit \cite{bonsall}.
For a typical electron density on a liquid-helium surface,
$n=10^{12}/{\rm m^{2}}$, $\omega_{0} \sim 4 \times 10^{10}
{\rm Hz}$.
The peak corresponds to the temporal oscillation in
the velocity autocorrelation first observed in
\cite{hansen,kalia}.
We have checked that the overall shape of the spectrum
does not change significantly with the sample size.

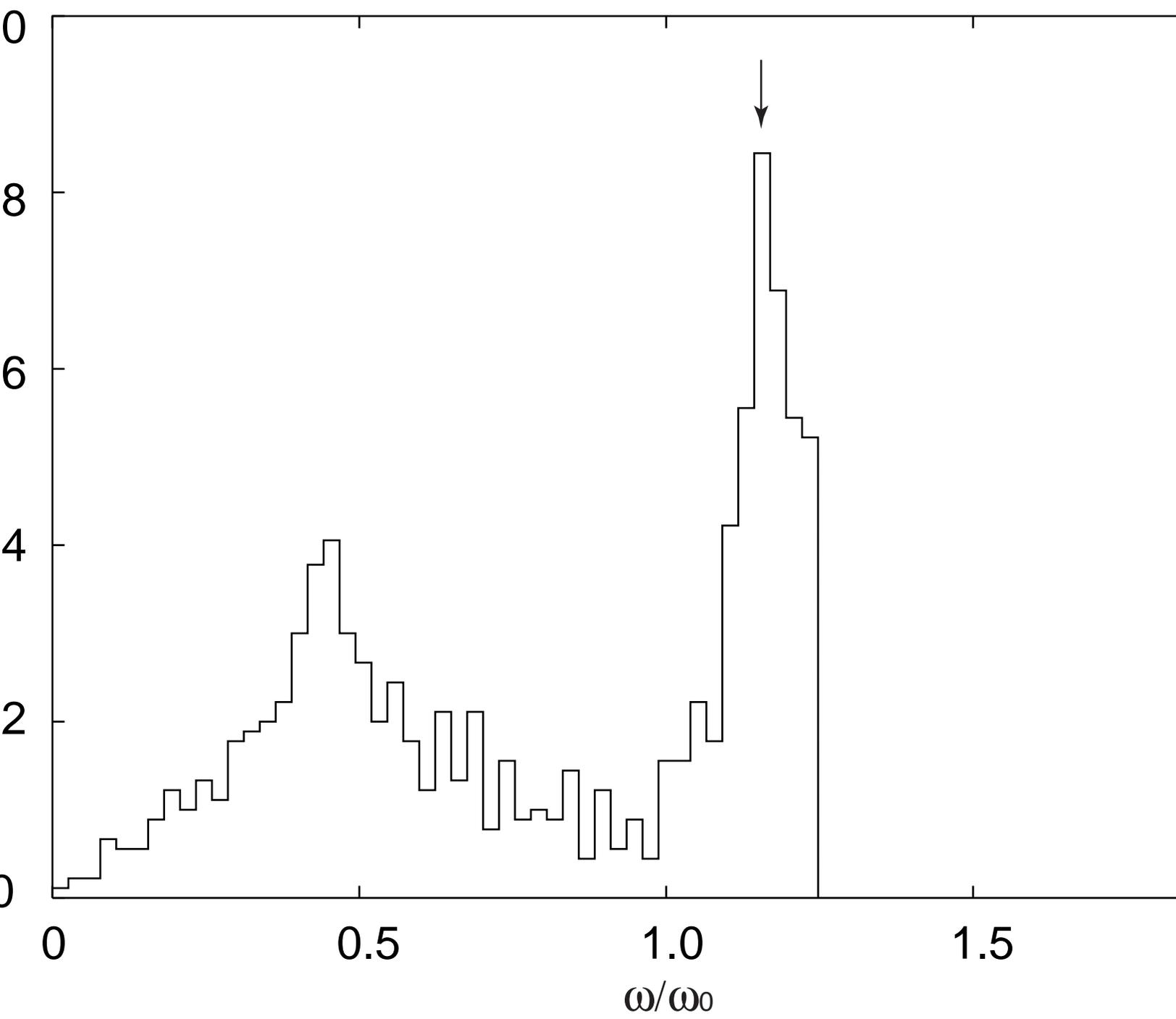
\begin{figure}
\caption{Vibrational density of states for the triangular
electron solid, obtained from an exact diagonalisation
of the dynamical matrix.}
\label{fig:dos}
\end{figure}

In order to identify the origin of the peak, we have
exactly diagonalised the dynamical matrix for the
finite-size triangular electron solid, from which
we obtain eigenfrequencies and eigenmodes.
We find that the vibrational density of states
(figure~\ref{fig:dos}) qualitatively reproduces
the velocity power spectrum for the solid phase.

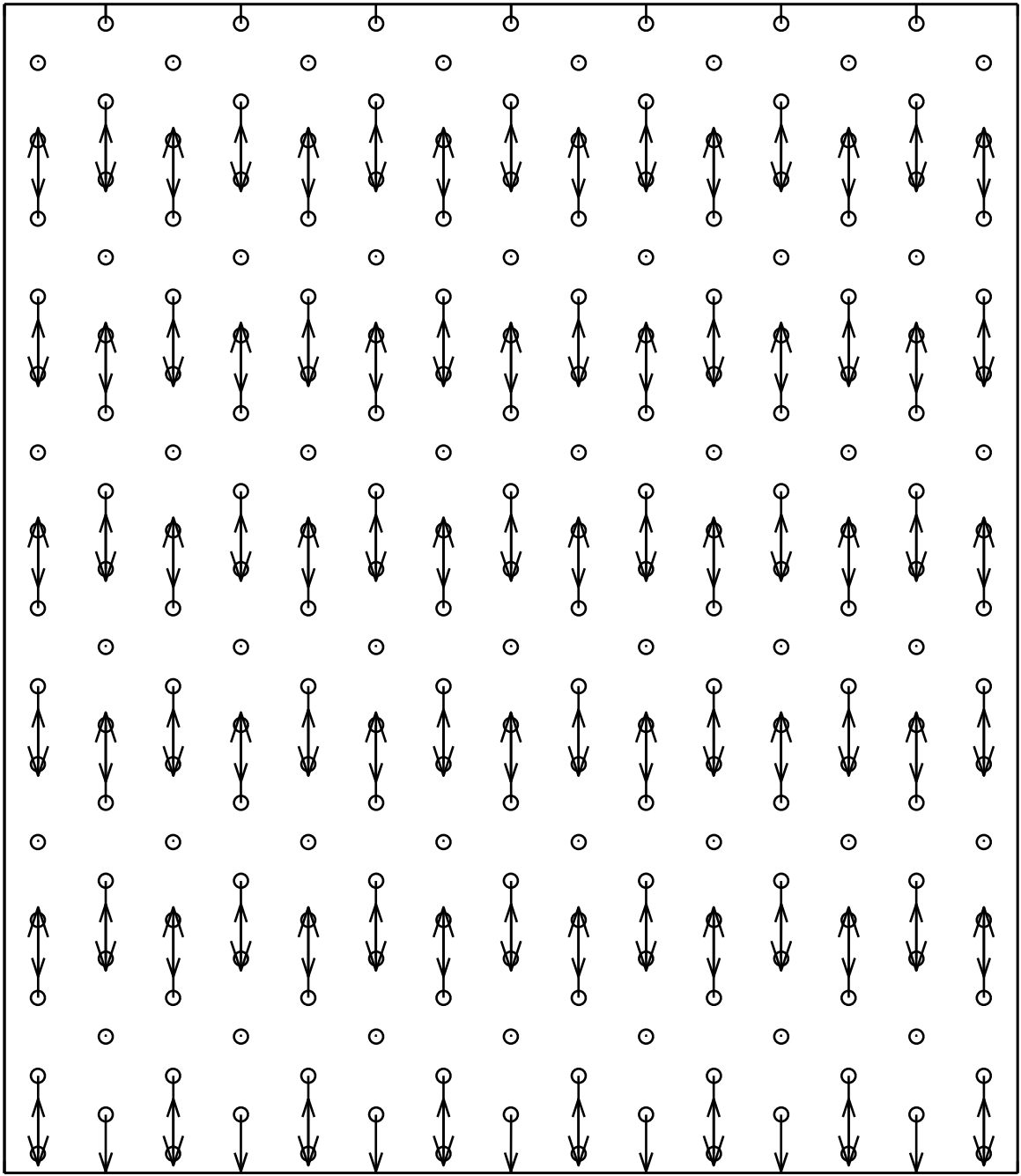
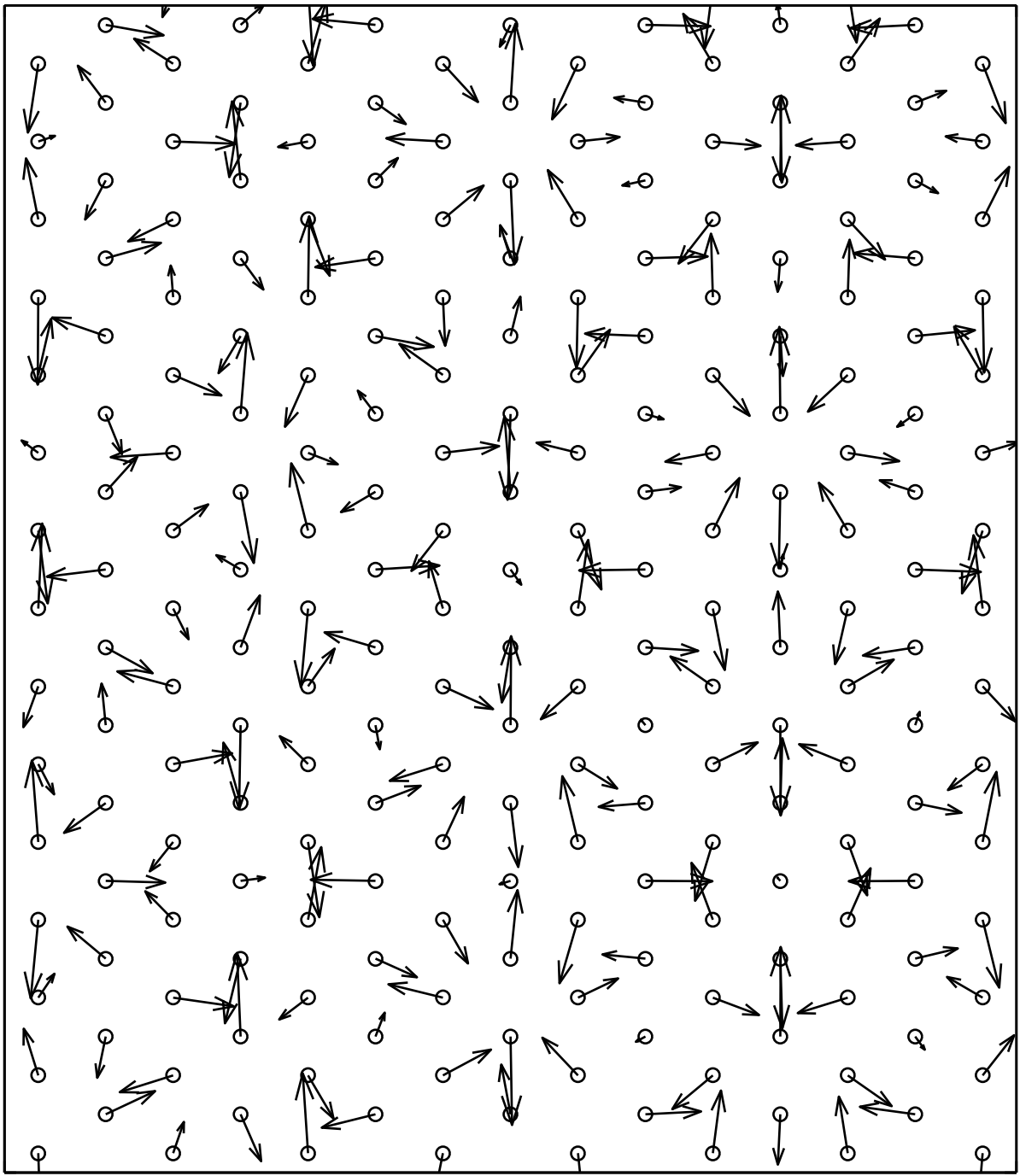
\begin{figure}
\caption{Typical vibrational modes around the high-frequency
peak (indicated by an arrow in figure~\protect\ref{fig:dos})
in the vibrational density of states.
(a) and (b) have almost the same frequency
[1.153 626 (a) and 1.156 830 (b) in units of $\omega_{0}$]
but different wave vectors.
A quarter of the triangular electron solid is shown
in either frame.}
\label{fig:mode}
\end{figure}

Typical vibrational modes around the high-frequency peak
turn out to be those which vibrate almost out of phase
between nearest-neighbour particles with the wavelength
$\sim d$, as are typically depicted in figure~\ref{fig:mode}.
The fact that the peak persists in the liquid phase is
considered to imply that the liquid has, despite the
absence of the long-range order, well-defined
local configurations that can sustain large wave-number
vibrations in finite spatial and temporal domains.

Our calculation is done for the long-ranged Coulomb potential.
It would be interesting to compare the result with those for
short-range potentials.

We would like to thank Professor Tsuneyoshi Nakayama for
valuable comments.
One of us (SM) wishes to thank Dr Katsunori Tagami for
discussions.
The computations were mainly done with Fujitsu VPP500 at
the Supercomputer Centre, Institute for Solid State Physics,
University of Tokyo.

\newpage

\end{document}